\documentclass[preprint2]{aastex}

\usepackage{graphicx}

\usepackage{natbib}
\bibpunct{(}{)}{;}{a}{}{,} 
\newcommand{\etal}{et al.}

\def \be {\begin{equation}}
\def \en {\end{equation}}

\def\gwt {G268556}
\def\gw {GW170104 }
\def\gwp {GW170104}
\def\t0{$10:11:58.599$}

\def \fv {}
\def \fvv {}
\def \fvvv {}
\def \fvr {}

\def \mt {}

\slugcomment{Submitted to \textit{Astrophysical Journal Letters}, May 31, 2017. Accepted, July 20, 2017.}

\shorttitle{AGILE search for e.m. counterpart of \gw}
\shortauthors{F.Verrecchia \etal}

\begin{document}

\title{AGILE Observations of the Gravitational Wave Source \gw}
%

\author{F.~Verrecchia\altaffilmark{1,2}, M.~Tavani\altaffilmark{3,4,5},
A.~Ursi\altaffilmark{3},
A.~Argan\altaffilmark{3}, C.~Pittori\altaffilmark{1,2},
I.~Donnarumma\altaffilmark{15}, A.~Bulgarelli\altaffilmark{6},
F.~Fuschino\altaffilmark{6}, C.~Labanti\altaffilmark{6}, M.~Marisaldi\altaffilmark{16}, Y.~Evangelista\altaffilmark{3},
G.~Minervini\altaffilmark{3},
A.~Giuliani\altaffilmark{7},
M.~Cardillo\altaffilmark{3}, F.~Longo\altaffilmark{9}, F.~Lucarelli\altaffilmark{1,2},
P.~Munar-Adrover\altaffilmark{17}, G.~Piano\altaffilmark{3},
M.~Pilia\altaffilmark{8}, V.~Fioretti\altaffilmark{6}, N.~Parmiggiani\altaffilmark{6},
A.~Trois\altaffilmark{8}, E.~Del~Monte\altaffilmark{3},
L.A.~Antonelli\altaffilmark{1,2},
G.~Barbiellini\altaffilmark{9},
P.~Caraveo\altaffilmark{7},
P.W.~Cattaneo\altaffilmark{10},
S.~Colafrancesco\altaffilmark{14},
E.~Costa\altaffilmark{3,15},
F. D'Amico\altaffilmark{15},
M.~Feroci\altaffilmark{3},
A.~Ferrari\altaffilmark{11}, A. Morselli\altaffilmark{12},
L.~Pacciani\altaffilmark{3},
F. Paoletti\altaffilmark{18,3}, A.~Pellizzoni\altaffilmark{8},
P.~Picozza\altaffilmark{12}, A.~Rappoldi\altaffilmark{10}, S.~Vercellone\altaffilmark{13}.
}

%
%

\affil {{\scriptsize $^{1}$ASI Space Science Data Center (SSDC), via del Politecnico, I-00133 Roma, Italy}}
\affil {{\scriptsize $^{2}$INAF-OAR, via Frascati 33, I-00078 Monte Porzio Catone (Roma), Italy}}
\affil {{\scriptsize $^{3}$INAF-IAPS, via del Fosso del Cavaliere 100,
I-00133 Roma, Italy}}

\affil {{\scriptsize $^{4}$Dipartimento di Fisica, Universit\`a di Roma ``Tor Vergata'',
via della Ricerca Scientifica 1, I-00133 Roma, Italy}}
\affil {{\scriptsize $^{5}$Gran Sasso Science Institute, viale Francesco
Crispi 7, I-67100 L'Aquila, Italy}}
\affil {{\scriptsize $^{6}$INAF-IASF-Bologna, via Gobetti 101, I-40129
Bologna, Italy}}

\affil {{\scriptsize $^{7}$INAF-IASF Milano, via E.Bassini 15, I-20133
Milano, Italy}}

\affil {{\scriptsize $^{8}$INAF, Osservatorio Astronomico di Cagliari,
via della Scienza 5, I-09047 Selargius (CA), Italy}}

\affil {{\scriptsize $^{9}$Dip. di Fisica, Universit\`a di Trieste and INFN,
via Valerio 2, I-34127 Trieste, Italy}}

\affil {{\scriptsize $^{10}$INFN-Pavia, via Bassi 6, I-27100 Pavia, Italy}}

\affil {
 {\scriptsize $^{11}$CIFS, c/o Physics Department, University of Turin, via P. Giuria 1,
 I-10125,  Torino, Italy}}

\affil {{\scriptsize $^{12}$INFN Roma Tor Vergata, via della Ricerca
Scientifica 1, I-00133 Roma, Italy}}
\affil {{\scriptsize $^{13}$INAF, Osservatorio Astronomico di Brera, via Emilio Bianchi 46, I-23807 Merate (LC), Italy}}
\affil {{\scriptsize $^{14}$University of Witwatersrand, Johannesburg, South Africa}}
\affil {{\scriptsize $^{15}$ASI, via del Politecnico snc, I-00133 Roma, Italy}}
\affil {{\scriptsize $^{16}$Birkeland Centre for Space Science, Department of Physics
and Technology, University of Bergen, Norway}}
\affil {{\scriptsize $^{17}$Unitat de F\'isica de les Radiacions, Departament de F\'isica,
and CERES-IEEC, Universitat Aut\`onoma de Barcelona, E-08193 Bellaterra, Spain }}
\affil {{\scriptsize $^{18}$East Windsor RSD, 25A Leshin Lane, Hightstown, NJ 08520, USA }}

\altaffiltext{1}{Email:francesco.verrecchia@ssdc.asi.it}

\vspace{-0.6cm}

\begin{abstract}

The LIGO/Virgo Collaboration (LVC) detected on 2017 January 4 a significant gravitational-wave (GW)
event (now named \gwp). We report in this Letter the main results obtained from the analysis of hard
X-ray and gamma-ray data of the AGILE mission that repeatedly observed the \gw localization region
(LR). At the LVC detection time $T_0$ AGILE observed about 36\% of the LR. The gamma-ray imaging
detector did not reveal any significant emission in the energy range 50 MeV--30 GeV. Furthermore, no
significant gamma-ray transients were detected in the LR that was repeatedly exposed over timescales
of minutes, hours, and days. We also searched for transient emission using data near $T_0$ of the
omnidirectional detector MCAL operating in the energy band 0.4--100 MeV. A refined analysis of MCAL
data shows the existence of a weak event (that we call ``E2'') with a signal-to-noise ratio of 4.4\,
$\sigma$ lasting about 32 ms and occurring $0.46$\,$\pm\,0.05 \,\rm s$ before $T_0$. A study of the
MCAL background and of the false-alarm rate of E2 leads to the determination of a post-trial significance
{\fvv between {\fvr 2.4$\,\sigma$} and 2.7$\,\sigma$} for a temporal coincidence with \gwp. We note that E2 has
characteristics similar to those detected from the weak precursor of GRB 090510. The candidate event
E2 is worth consideration for simultaneous detection by other satellites. If associated with
GW170104, it shows emission in the MeV band of a short burst preceding the final coalescence by 0.46
s and involving $\sim 10^{-7}$ of the total rest mass energy of the system.

\end{abstract}

\keywords{gravitational waves, gamma rays: general.}

    \section{Introduction}


The LIGO/Virgo Collaboration  (LVC) detected a
gravitational-wave (GW) event on  2017 January 4 \cite[][]{2017GCN..20364...1S,2017GCN..21056...1B,2017GCN..20385...1G}.
 This event, originally labeled \gwt, occurred at time
$T_0$\,=\,10:11:$58.599$ UTC; it is now named  \gw \cite[][ hereafter A17]{2017PhRvLsub}.
The event is identified
as a compact binary coalescence (CBC) of two black holes of masses
$ \sim 31 \, M_{\odot}$ and $\sim 19 \, M_{\odot}$. The estimated
distance is $880^{+450}_{-390}\, \rm Mpc$ corresponding to a redshift
of $z\,=\,0.18^{+0.08}_{-0.07}$.
%
\gw is an event of great interest because  its ``false-alarm rate''
(FAR) is less than 1 in 70,000 years as determined by a refined
analysis (A17). 
%
This event is the third of a set of high-significance and confirmed GW events detected by
LVC, the first one being GW150914 and the second one being GW151226 \cite[][]{2016PhRvD..93l2003A,2016PhRvL.116m1103A,2016PhRvL.116x1103A,2016PhRvL.116f1102A,2016PhRvD..93l2004A,2016PhRvL.116x1102A}.

 A first sky map of \gw was distributed through a LVC-GCN on 2017 January 4th \cite[][]{2017GCN..20364...1S},
including an initial localization generated by the BAYESTAR pipeline \cite[][]{2016PhRvD..93b4013S}.
An updated sky map was distributed on Jan. 17, 2017 \cite[][]{2017GCN..20385...1G},
based on LALInference \cite[][]{2015PhRvD..91d2003V}.
 The probability is concentrated in two long, thin arcs in the sky.
The 50\% credible region spans about 500 $\rm deg^2$ and the 90\% region about 2000 $\rm deg^2$.
A final sky map was issued on 2017 May 17 \cite[][]{2017GCN..21056...1B}.
%
 AGILE reacted to the initial LVC
notification of \gwp, and started a quicklook analysis within a few hours
as {\fv discuss}ed below \cite[][]{2017GCN..20375...1T,2017GCN..20395...1T}.
In the following, we
name differently the two arcs of the \gw LR differently: the northern arc is
named ``A'', and the southern arc is named ``B''.

In this Letter we present the main results of the analysis of AGILE
data concerning \gwp. Sect.~2 summarizes the observation
capability of AGILE for the search of GW event counterparts. Sect.~3
presents the results concerning gamma-ray emission above 50 MeV
from \gwp. Sect.~4 includes the analysis of gamma-ray data
obtained by the MCAL detector in the energy range 0.4-100 MeV with
a discussion of an interesting candidate event occurring near
$T_0$. Sect. 5 presents a brief discussion of our results and
conclusions {\fv are reported in Sect.~6}.



\section{The AGILE satellite searching for GW event counterparts}

The AGILE satellite that is orbiting the Earth in a near
equatorial orbit at an altitude of $\sim$\,500 km
\cite[][]{2009A&A...502..995T} is currently exposing 80\% of the
entire sky every 7 minutes in a ``spinning mode''.
 The instrument consists of an imaging gamma-ray Silicon
Tracker (sensitive in the energy range 30 MeV--30 GeV),
Super-AGILE (SA; operating in the energy range 20--60 keV), and the
Mini-Calorimeter \citep[MCAL; working in the range 0.35--100
MeV;][]{2009NIMPA.598..470L,2008A&A...490.1151M,2008NIMPA.588...17F}
with an omnidirectional field of view (FoV) and self-triggering capability
in burst-mode for various trigger timescales. The
anticoincidence (AC) system completes the instrument \cite[for a
summary of the AGILE mission features, see ][]{2009A&A...502..995T}.
The combination of Tracker, MCAL, and AC working as a gamma-ray
imager constitutes the AGILE-GRID. The instrument is capable of
detecting gamma-ray transients and GRB-like phenomena {\fv on}
timescales ranging from submilliseconds to tens of hundreds of
seconds.
 In preparation of the Advanced LIGO second Observing run (O2) and the
first Advanced Virgo one, the AGILE data acquisition system and
trigger configuration were substantially improved. The data
obtained for \gw greatly benefited from this improvement.
%
%

Both MCAL and SA are particularly efficient in detecting
GRBs of the long and short classes as demonstrated in recent
years\footnote[2]{GRB 080514B, \cite[][]{2008A&A...491L..25G},
GRB 090401B, \cite[][]{2009GCN..9069....1M}, GRB 090510,
\cite[][]{2010ApJ...708L..84G}, GRB 100724B,
\cite[][]{2011A&A...535A.120D}, an evaluation of GRID upper limits
for a sample of GRB in \cite{2012A&A...547A..95L}, GRB 130327B,
\cite[][]{2013GCN..14344...1L}, GRB 130427A,
\cite[][]{2013GCN..14515...1V}, GRB 131108A,
\cite[][]{2013GCN..15479...1G,2014arXiv1407.0238G}, and the first catalog of GRB detected by MCAL \cite[][]{2013A&A...553A..33G}.}.
%
 Because of its triggering capabilities, in particular a special
submillisecond search for transient events detected by MCAL
\cite[][]{2009NIMPA.598..470L,2009A&A...502..995T}, AGILE so far has further detected 
more than 1000 terrestrial gamma-ray flashes with
durations ranging from hundreds to thousands of microseconds
\cite[][]{2014JGRA..119.1337M,2011PhRvL.106a8501T}.
%
%

The characteristics that make AGILE in spinning mode an important
instrument for follow-up observations of wide GW source LRs are: a
very large FoV of the GRID (2.5 sr), an accessible region of 80\%
of the whole sky that can be exposed every 7 minutes,
 100--150 useful passes every day for any region in the accessible sky, a gamma-ray exposure
of $\sim$\,2 minutes of any field in the accessible sky every 7 min, sensitivity of
$\sim 10^{-8} \, \rm erg \, cm^{-2} \, s^{-1}$ above 30 MeV for typical single-pass
of unocculted sky regions, a submillisecond MCAL trigger for very fast events in
the range 0.4--100 MeV, and hard X-ray (20--60 keV) triggers of GRB-like events with
a localization accuracy of 2--3 arcmin in the SA FoV ($\sim 1$ sr) when operating in
imaging mode.


Satellite data are transmitted to the ground currently on average
every passage over the {\mt ASI} Malindi ground station in Kenya.
Scientific data are then processed by {\mt a } fast processing
{\mt that was recently} improved for the {\mt search of
electromagnetic (e.m.) counterparts of GW sources}. The current
data processing can typically produce an alert for a transient
gamma-ray source and/or GRB-like events within 20 min--2 hrs  from
satellite onboard acquisition \cite[][]{2014ApJ...781...19B,2013NuPhS.239..104P}.

\section{GRID observations of \gw}

\subsection{Prompt emission}

\begin{figure}[t]
   \centerline{\includegraphics[width=9.0cm, angle = 0]{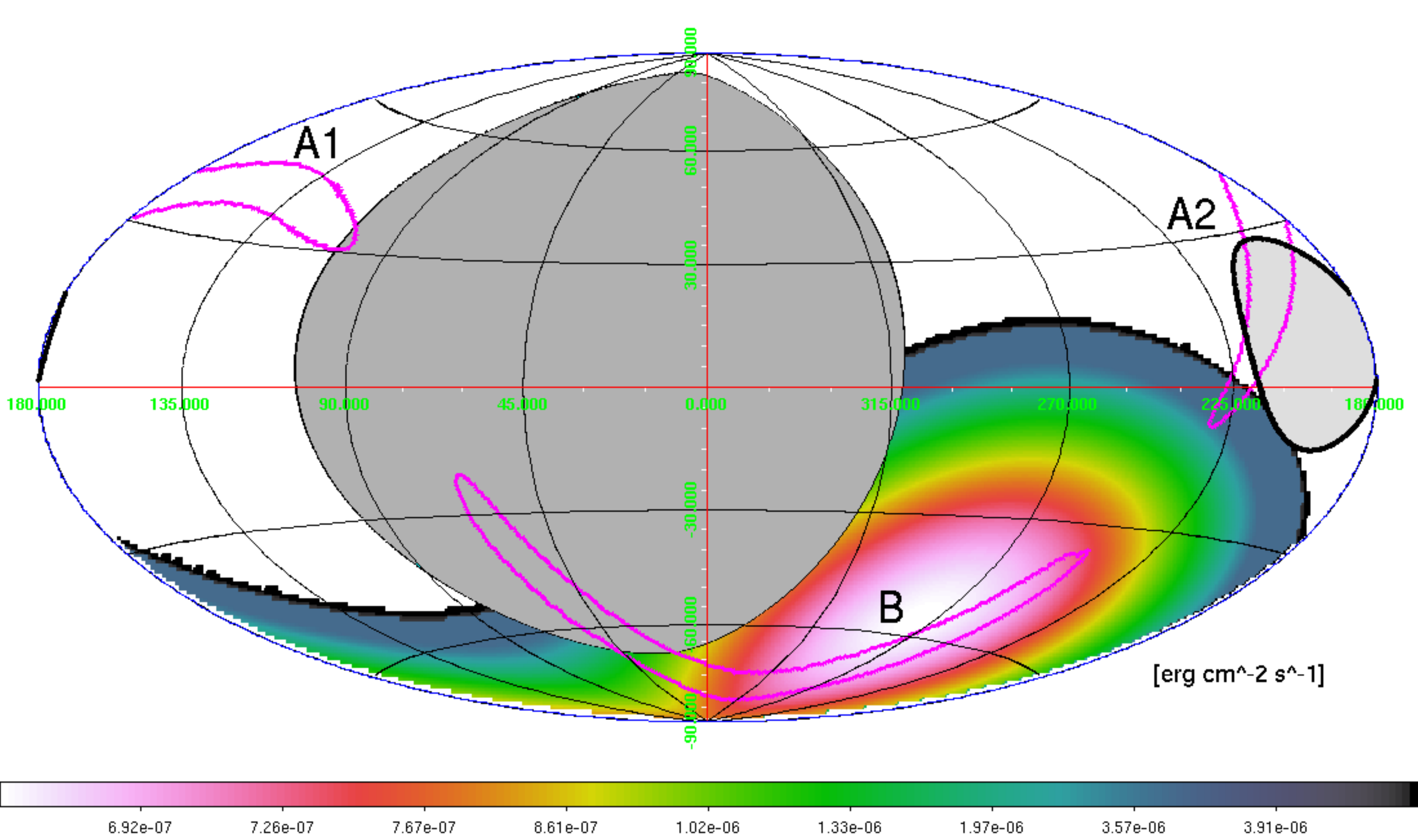}}
    \caption{AGILE-GRID $E > 50$ MeV sensitivity map {\fv (in $\rm erg\, cm^{-2} s^{-1}$)} in Galactic coordinates based on the
     gamma-ray exposure at the detection time $T_0$ of \gwp.
The shadowed areas show the Earth-occulted region and the sky fraction not directly accessible by
AGILE for solar panel constraints. The magenta contours show the {\fv LALinference} 90\% {\fv c.l. LR} of \gw \cite[][]{2017GCN..20385...1G}.
    The two parts of the northern arc of the LR
 are marked with ``A1'' and ``A2'' while the southern arc is marked by ``B''.
The AGILE instrument has significant exposure of the B-region at
$T_0$, covering about 35\% of the total LR of \gwp.}
 \label{fig-1}
 \end{figure}

AGILE sky scanning observation covered the \gw event \cite[at time
$T_0$\,=\,10:11:58.599 UTC on 2017 January 4;][]{2017GCN..20364...1S}
with the GRID FoV only marginally
occulted by the Earth. Fig.~\ref{fig-1} shows the gamma-ray
{\fv sensitivity map} above 50 MeV at the \gw {\fv detection} time.
 The Earth partially
covers the lower part of the B arc (obtained from the 90\% \gw LR
from the refined localization map; A17).
The remaining part of the B arc and a small area of the upper
A arc (A2) are not occulted and inside the AGILE exposure region.
We performed a search for (1) the prompt event involving the
GRID, MCAL, AC and SA and
(2) precursor and delayed emission on multiple timescales
involving the GRID.
%
%
{\fv Interestingly}, the AGILE-GRID exposed a good fraction (36\%) of the
B arc on time interval of about 200 s centered at $T_0$.
%

\begin{figure}[t]
    \centerline{\includegraphics[width=9cm, angle = 0]{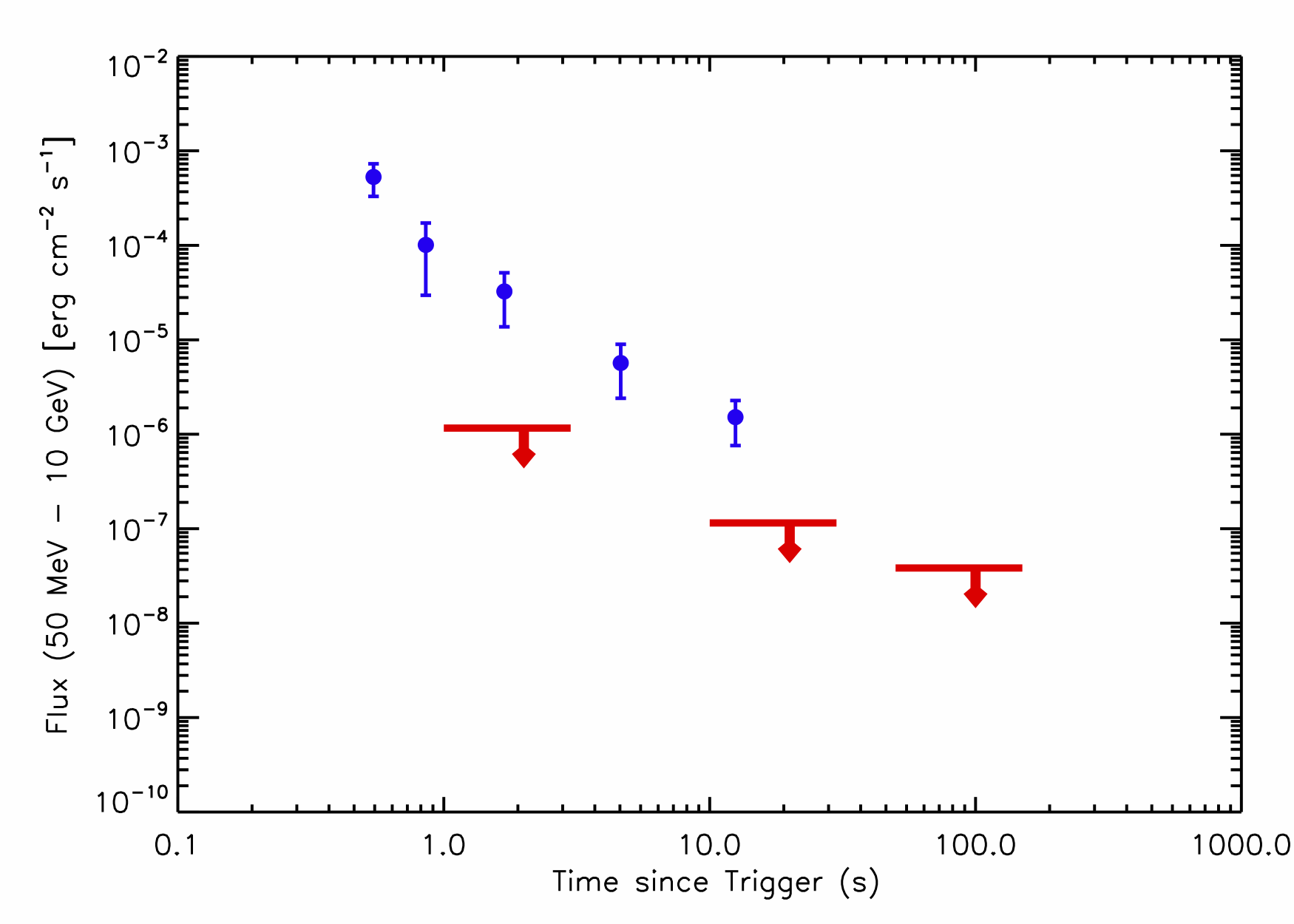}}
    \caption{Gamma-ray light curve obtained by the AGILE-GRID for the short GRB 090510 repositioned at
redshift $z = 0.18$ of \gw  (blue points).
    The GRID upper limits obtained for the exposed fraction of
the B arc of \gw at different timescales are marked in red. }
 \label{fig-3}
 \end{figure}

No significant gamma-ray emission was detected by the GRID near $T_0$.
 We obtained $3\,\sigma$ flux upper limits (ULs) in the band 50 MeV--10
GeV, $UL = 1.8 \times 10^{-6} \rm \, erg \, cm^{-2} \, s^{-1}$ and
$UL = 2.9 \times 10^{-8} \rm \, erg \, cm^{-2} \, s^{-1}$ for 2
s and 100 s integrations, {\mt respectively}.
As shown in Fig.\ref{fig-3}, we compare our gamma-ray ULs
 for \gw with the short GRB 090510 gamma-ray light curve detected by AGILE
\cite[][]{2010ApJ...708L..84G}. The short GRB 090510 (at the redshift $ z = 0.9$)
has been considered a reference for {\fv potential} electromagnetic gamma-ray emission
that could be associated to a GW event
\cite[][]{2016ApJ...823L...2A,2016ApJ...825L...4T}.
In Fig.~\ref{fig-3} we show the GRB 090510 light curve
rescaled (in flux and time corrected) as if it were emitted at the
\gw estimated distance ($z=0.18$).

\begin{table}[t!]
\begin{center}
{\bf Table 1: Analysis of individual passes over the \gw
localization region} \vskip.3cm
\begin{tabular}{|c|c|c|c|c|c|}
  \hline
 Expo- & Bin   &  3\,$\sigma$ UL   & \multicolumn{3}{c|}{Fraction$^{c}$} \\
 sure   & Central   & Range$^{b}$  & \multicolumn{3}{c|}{ } \\
 N$^o$  & Time$^{a}$ &  & \multicolumn{1}{l|}{Total}&\multicolumn{1}{l|}{ B }&\multicolumn{1}{l|}{ A } \\
  \hline

  -10& -950 &  2.9/84.0  & 50\% & 100\% & 20\%  \\
  -9 & -850 &  4.1/105.0 & 72\% & 100\% & 10\% \\
  -8 & -750 &  2.9/95.0  & 39\% &  22\% & 53\%   \\
  -7 & -650 &  11.0/59.0 & 34\% &   0\% & 62\%   \\
  -6 & -550 &  8.5/108.0 & 33\% &  62\% & 10\%  \\
  -5 & -450 &  2.9/74.0  & 50\% & 100\% & 9\%   \\
  -4 & -350 &  4.1/91.0  & 63\% & 100\% & 36\%  \\
  -3 & -250 &  2.9/103.0 & 36\% &  20\% & 58\% \\
  -2 & -150 &  7.9/51.0  & 32\% &   0\% & 58\%   \\
  -1 & -50  &  11.0/89.0 & 30\% &  56\% & 9\%  \\
   0 &  0   &  2.9/110.0 & 36\% &  64\% & 9\%  \\
  +1 & +50  &  4.5/98.0  & 43\% &  56\% & 33\% \\
  +2 & +150 &  3.0/66.0  & 32\% &   0\% & 28\%  \\
  +3 & +250 &  5.9/42.0  & 32\% &   0\% & 29\%  \\
  +4 & +350 &  17.0/105.0& 25\% &  19\% & 5\%  \\
  +5 & +450 &  3.0/42.0  & 25\% &   8\% & 5\%  \\
  +6 & +550 &  4.8/110.0 & 29\% &  17\% & 12\% \\
  +7 & +650 &  3.2/73.0  & 30\% &  29\% & 1\%    \\
  +8 & +750 &  4.8/33.0  & 31\% &  29\% & 2\%  \\
  +9 & +850 &  34.0/108.0& 15\% &  10\% & 5\% \\
  +10& +950 &   3.2/54.0 & 15\% &  10\% & 5\%   \\
  \hline
\end{tabular}
\end{center}
\noindent {\bf \scriptsize Notes.}
\tablenotetext{a}{\scriptsize Centroid of the window time interval with respect to 
$T_0$, $t- T_0$,\\ in seconds. All integrations have a duration of 100 s.}
\tablenotetext{b}{\scriptsize $3\,\sigma$ flux upper limit ($10^{-8} \, \rm erg \, cm^{-2} \, s^{-1}$) obtained for emission in \\the range 50 MeV--10 GeV and for a spectrum assumed to be\\ similar to the delayed gamma-ray emission of the short \\GRB~090510.}
\tablenotetext{c}{\scriptsize Fraction of the GRID exposed whole LVC LR of \gw\\ and of the B and A arcs respectively.}

\end{table}

\subsection{Precursor and Delayed Emission}

\begin{figure*}  [t!]
   \centerline{\includegraphics[width=\textwidth]{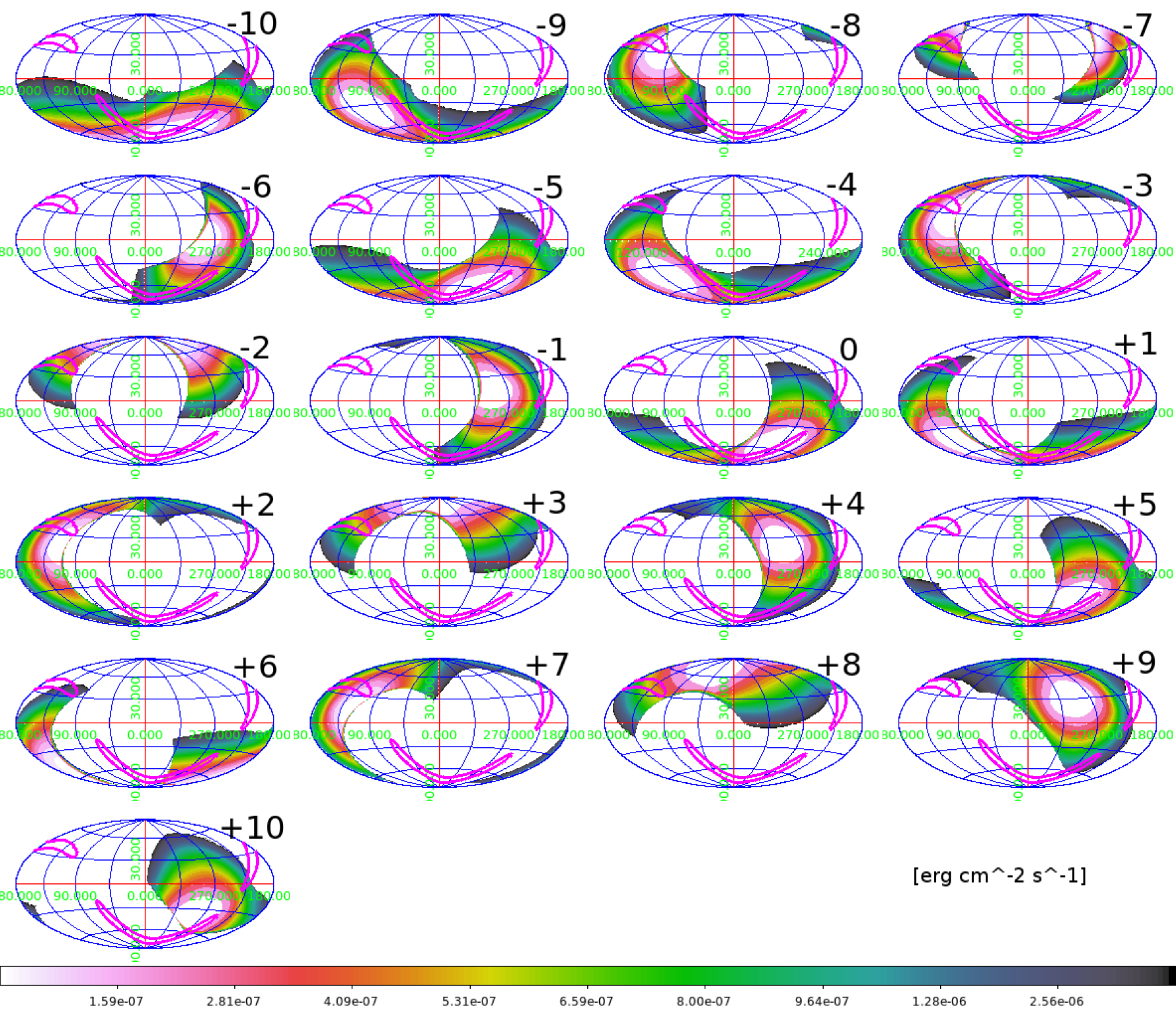}}
\caption{Sequence of ($E > 50$ MeV)
maps in Galactic coordinates showing the AGILE-GRID passes with
the best sensitivity {\fv (in $\rm erg\, cm^{-2} s^{-1}$)} over the \gw LR
obtained during the period (- 1000 s, + 1000 s) with respect to
$T_0$. The color maps show the gamma-ray flux 3\,$\sigma$
upper limits in the range 50 MeV--10 GeV with the most
stringent values being $ UL = (1-2) \times 10^{-8} \rm \, erg \,
cm^{-2} \, s^{-1}$. The sequence shows 21 maps for all the one-orbit
passes of Table 1, corresponding to the 100 s interval numbers
(from top left to bottom right): from $-10$ to $+10$. The \gw
LR is marked by the purple contour, the LALinference 90\% {\fv c.l.} \cite[][]{2017GCN..20385...1G}.
} \label{fig-2}
\end{figure*}

AGILE was also optimally positioned in the \gw LR at interesting time intervals
preceding and following the prompt event.
%
We show in Fig. \ref{fig-2} the sequence of passes of 100 s
duration over the LR within $T_0 \pm \,1000$ s, and we summarize in
Table 1 the 3\,$\sigma$ flux UL range, minimum and maximum value
within the covered LR portion. We label time intervals of the highest
GRID exposure with progressive numbers with respect to the prompt
interval $\Delta_0$.
It is interesting to note that exposures obtained from Table 1
reached 60-70\% of the LR.
%
%
We also carried out a long-timescale search for transient
gamma-ray emission during the hours immediately following or
anticipating the prompt event. No significant gamma-ray emission
in the \gw LR was detected over timescales of hours up to two-day integrations.

\begin{figure*}
    \centerline{\includegraphics[width=\textwidth, angle = 0]{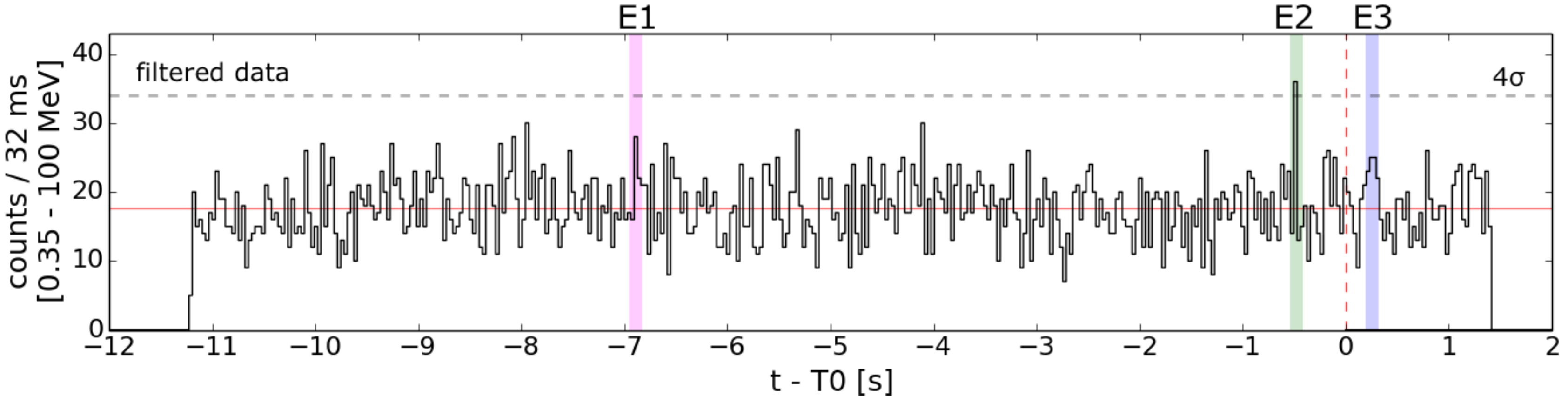}}
    \caption{Light curve of MCAL data which includes the \gw event time. Data are displayed with a 32 ms time binning for the MCAL full energy band (0.35--100 MeV), after refined data processing.
The E1, E2 and E3 event times are marked by vertical magenta, green and light blue lines, while the $T_0$ is marked by a dashed red line.
The horizontal dashed grey line indicates the 4\,$\sigma$ level, estimated on the whole
data acquisition interval (12.6 s). The orange horizontal line marks the average background level.
    }
 \label{fig-4}
\end{figure*}

\section{Analysis of MCAL data and identification of an interesting event}

The onboard MCAL processing procedure identified an
above-threshold event that triggered full telemetry data
acquisition for a 12.6 s segment starting at time $T_{1}$
= 10:11:47.4 UTC and ending at $T_2$ = 10:12:00.0 UTC.
{\fv This event occurred on the 16 ms trigger timescale for which the
average MCAL occurrence rate\footnote[3]{This rate is dominated by background events. An analysis of the triggered MCAL data with 16 ms bin
shows that the rate of data acquisitions with peaks of signal-to-noise (S/N) above 5\,$\sigma$ is $\sim\,5\,\times 10^{-4} \,\rm Hz$.} is 7.5\,$\times 10^{-3} \rm Hz$.}
Interestingly, this time interval includes the \gw {\fv detection} time
$T_0$.
%
%
 Analysis in this time interval reveals a significant
peak at the onboard trigger time $T_{E1} = T_0 - 6.87 \rm \, s$ (10:11:51.70\, $\pm 0.05
\, \rm s$ UTC, that we call ``E1'').
This event {\fv {\fvvv appears to be significant with} 26 ms binning}
{\fv resulting in} 37 raw counts in the full {\fv energy} band
and a preliminary signal-to-noise (S/N) ratio of 5.2\,$\sigma$.
After having carried out standard off-line processing and event
pile-up removal, the E1 S/N ratio becomes $3.6 \, \sigma$.
%
%
Furthermore, two more events within 1 s from $T_0$ are 
{\fv noticed} in the raw telemetry for the full MCAL energy band.
{\fv An event} (that we call ``E2'') occurred at
$T_{E2} = T_0 - 0.46 \, \rm s$ (10:11:58.10\, $\pm 0.05 \,
\rm s$ UTC) {\fv and {\fvvv is significant with} 32 ms binning. This
event shows 37 raw counts and pre-filtering {\fvvv S/N} ratio
of 4.2\,$\sigma$. After filtering and pile-up removal its S/N ratio
slightly increases to 4.4\,$\sigma$ (see Fig.~\ref{fig-4}). A third feature after $T_0$
(that we call ``E3'') is noticed in the 96 ms binned light curve
occurring on $T_{E3} = T_0 + 0.30 \, \rm s$
(10:11:58.90\, $\pm 0.05 \, \rm s$ UTC). This feature shows 89 raw
counts and a pre-filtering {\fvvv S/N} ratio
of 3.7\,$\sigma$. After filtering and pile-up removal its S/N
ratio is 3\,$\sigma$(in Figure 5 a 4s  interval centered at E2 time of the 32 ms light curve, in the full band and in the E\,$\leq$\,1.4 MeV, E\,$\geq$\,1.4 MeV energy bands).}

{\fvvv Therefore, with these preliminary indications, 
we focused on detecting signals in specific time bins that turn out to be the most effective ones
in revealing short GRBs as detected by the AGILE-MCAL. We performed a refined 
search {\fvr (with time shifts)} in three binning time scales (26, 32, and 96 ms). 
One relevant timescale is of the order of 100 ms as demonstrated by the
range of T$_{50}$ measured in MCAL data \cite[see Fig.~\ref{fig-4} of
][]{2013A&A...553A..33G}. This timescale is optimal for searching
emission during the prompt phase of short GRBs (e.g.,
GRB~090510, and more recently GRB~170127C, \citealt{2017GCN..20545...1U}). The
other two timescales (26 and 32 ms) are chosen to optimize
searches of "GRB~090510-precursor-like events" known to last
between 20 and 30 ms. The latter class of events constitutes the
physical reference of our optimized onboard trigger logic for
data acquisition that AGILE implemented since November 2016 to
improve its sensitivity to GW events.}
  We applied to each binned light curve
different phase shifts: 4 equally spaced shifts for the 32 and 96
ms binned light curves and 2 shifts for the 26 ms one,
corresponding to shifts of $1/4$ and $1/2$ of the time bins,
respectively.
%
{\fv These events/ features are {\fvv presented} in the 32 ms binned
light curve of Fig.~\ref{fig-4}
showing filtered and pile-up removed data}. {\fv We verified {\fvv that} the
South Atlantic Anomaly passage occurs $\sim \,1700\, \rm s$ after $T_0$}.

\begin{figure} [t!]
 \vspace*{0.2cm}
   \centerline{\includegraphics[width=9.0cm]{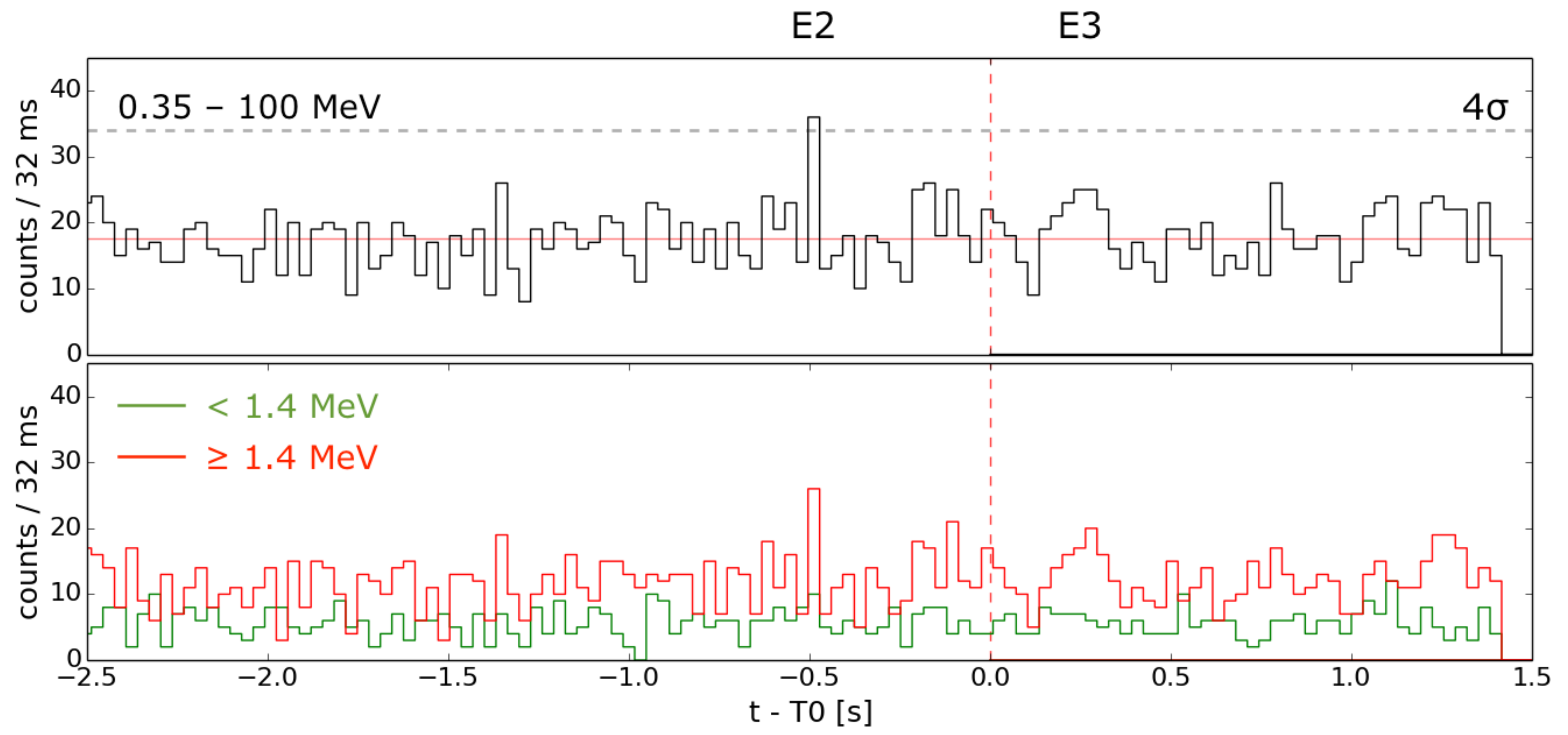}}
\caption{ MCAL light curve binned at 32 ms for the 4 s time interval centered on the E2 event. Data displayed
 for the MCAL full energy band (0.35 - 100 MeV, top panel), and for the soft ($\leq 1.4 \, \rm MeV$) and hard ($\geq 1.4 \, \rm MeV$)
energy bands (bottom panel).
 The \gw event time is at the origin of the abscissa.
In the top panel the horizontal dashed grey line indicates the 4\,$\sigma$ level, estimated on the whole
data acquisition interval (12.6 s). The orange horizontal line marks the average background level.
}
\label{fig-5}
\end{figure}

In order to assess the post-trial probability of detecting these
three events close to $T_0$ we developed a targeted procedure to
identify untriggered peaks with the characteristics of E1, E2, and E3, {\fvv within} their respective
light curves (of binning equal to 26, 32, and 96 ms).
This procedure was carried out over a time interval of 2 weeks centered at $T_0${\fvv , for a total livetime of 104,557 s}.
Given the MCAL filtered telemetry, this analysis produces FARs for these events
and specific choices of time binning. The results\footnote[4]{{\fvvv Our signal is a superposition of Poisson noise and sparse non-Poissonian events. We conservatively ignore the subtraction of the latter events and determine the FAR as deduced from the total signal.}} are:
 $FAR_{E1} = 1\,\times\,10^{-3}$ Hz for 26 ms binning, $FAR_{E2} = 1\,\times\,10^{-4}$
 Hz for 32 ms binning, and $FAR_{E3} = 3\,\times\,10^{-3}$ Hz for 96 ms binning.
{\fvv Only e}vent E2 {\fv potentially} emerges as an interesting candidate; further
analysis supports this conclusion. {\fvvv It is then important to obtain the FARs for the 26 and 96 ms binned light curves corresponding to detections with the same S/N of E2. We obtained ${\fvvv 1.6\,\times\, 10^{-4}}$ and ${\fvvv 5.7\,\times\, 10^{-5}}$ Hz for the 26 and 96 ms binned light curves, respectively.}
%
{\fvvv We take ${\fvvv FAR_{E2} = 1\,\times\,10^{-4}}$ Hz as representative of our search of events similar to E2.} {\fvr In addition to the refined search binnings, for completeness in the analysis of the relevant time scales in the number of trials calculation, we also include the 16 and 64 ms unshifted binnings that refer to the onboard trigger logic time scales initially used for our data processing. We determine the FARs with the same S/N of E2 to be ${\fvr 2.4\,\times\, 10^{-4}}$ and ${\fvr 7.5\,\times\, 10^{-5}}$ Hz, respectively.}

Having obtained the FAR over a 2-week timescale, we then estimate the post-trial probability $P$ of these events accidentally occurring within a time interval
$\delta \, t$ from $T_0$ {\fvv with two different methods. The first one follows} the formalism and definition laid out by \cite{2016ApJ...826L...6C} for the analysis of Fermi-GBM data of GW150914.
{\fvv In that work, t}he probability {\fvv is defined as}
$P\,= \, N \times\, FAR \,\times\, \delta \, t \,\times (1 + \ln(\Delta t/ t_{bin}))$, where
$N$ is the number of trials,
$FAR$ is the false-alarm rate, $\delta \,t$ is the time difference between the event time and $T_0$,
$\Delta \, t$ is the one-sided time interval over which the search is done, and
$t_{bin}$ is the time binning used in the analysis.
We {\fvv considered as an event of interest only E2 for which we}
obtained:
$P^{E2} =\,1{\fvr 2} \times \, (1 \times 10^{-4} \, {\rm Hz}) \,\times
(0.46 \, {\rm s}) \times (2 + \ln(11.2 \, {\rm s}/ 0.0{\fvr 1}6 \, {\rm
s}) + \ln(1.4 \, {\rm s}/ 0.0{\fvr 1}6 \, {\rm s})) \simeq 0.00{\fvr 719}\, (\sim  2.4 \,\sigma)$,
%
%
{\fvv where we included both the backward and forward time intervals $\Delta t$ of 11.2 and 1.4 $\rm s$ respectively, and $\delta \, t = 0.46 \, \rm s$. We adopted a factor N$\,=\,(4 \times 2) + (2 \times 1){\fvr + 1 + 1}\,=\,12$ to take into account
the number of trials implemented in the refined analysis procedure, which include{\fvvv s} three independent searches
and their corresponding time shifts. {\fvvv T}his number of trials is overestimated {\fvvv because} the
results for overlapping phase shifts are not independent {\fvvv and for the effect of unsubtracted non-Poissonian events in the FAR calculation}.}

{\fvv The second method to estimate the post-trial probability of E2 is based on an independent procedure. We considered the whole MCAL data stream of the same two weeks used for FAR estimate {\fvvv with "a-posteriori" choices: 32 ms binning,} and fixed time shift. We simulated the occurrence of an external trigger, at a random time $t^{*}$, and determined the time difference ($\tau$) between $t^{*}$ and the closest peak with S/N\,$\geq\,$4.4\,$\sigma$. This procedure was repeated 6 million times and produced a distribution of $\tau$'s. We determine the probability of occurrence of a signal with characteristics similar to E2 ($\tau\,\leq \,$0.46 s) to be 3.7\,$\times\,10^{-3}$, that corresponds to 2.7\,$\sigma$. We consider this value to be an alternative estimate of the post-trial probability {\fvvv of E2 events}. 

{\fvv We conclude that the} E2 event turns out to {\fvv have} a post-trial {\fvv coincidence} probability {\fvv between {\fvr 2.4} and 2.7$ \, \sigma$}.

\section{Discussion}

AGILE observed the field containing \gw with good coverage and
significant gamma-ray exposure of its LR. The AGILE-GRID {\mt
exposed 36\% of the} B arc at the detection time $T_0$. No transient
gamma-ray source was detected near $T_0$ over timescales of 2, 20,
and 200 s starting at $T_0$. Fig. 2 shows these prompt upper
limits compared to a rescaled gamma-ray light curve of GRB~090510.
{\fv Our gamma-ray upper limits are quite relevant in constraining
prompt gamma-ray emission from \gwp.} Furthermore, we obtain
constraints for gamma-ray emission above 50 MeV also for possible
precursor and delayed emission as detailed in Table 1.
%
%
It is important to note that the co-axial SA detector had a partial
coverage of the \gw LR at $T_0$ (within 1s), that was observed
between 0 and 30$^{\circ}$ off-axis. No significant detection was
obtained by SA imaging or ratemeter data
in the 20--60 keV band during the time interval $T_0 \pm$\,100 s
for a stable background. The $3\,\sigma$ UL has been derived for a 1 
s integration time and varies between $F^{SAc}\,=\, 1.5 \times 10^{-8} \, \rm erg \,
cm^{-2}$ for an on-axis position and $F^{SAs}\,=\, 6.6 \times 10^{-8} \, \rm erg \,
cm^{-2}$ at 30$^{\circ}$ off-axis position. A search in the SA
ratemeter data does not produce a significant detection with a
$2\,\sigma$ fluence upper limit of $2.4 \times 10^{-8} \, \rm erg \,
cm^{-2}$ {\fv (1 s integration time)}.

MCAL detected three short-timescale events/features, of which only
one remained significant after a refined analysis. The E2 event has a pre-trial
S/N ratio of 4.4\,$\sigma$, and a post-trial probability of occurring within
0.46 s before $T_0$ {\fv between {\fvr 2.4} and 2.7\,$\sigma$}.
This event is the most interesting and is
worthy of additional investigation and searches for possible
coincident detections by other space instruments. {\fvv We estimated t}he observed
energy flux in the energy range 0.4--40 MeV {\fvv at two positions on the B arc, and
at one on the A1 arc. We used a mean photon index taking into account the detector
response and fitted the normalization parameter.
We obtained on the B arc
$ F_{Bc}\,=\,2.8_{-1.0}^{+1.9} \times 10^{-6} \, \rm erg \,  cm^{-2} \, s^{-1}$ {\fv assuming
a power-law model with a fixed mean photon index of -2 and estimating it at the a position with the best sensitivity (on-axis)},
while F$_{Bs}\,=\,2.5_{-1.0}^{+2.7} \times 10^{-6} \, \rm erg \,  cm^{-2} \, s^{-1}$
with the same photon index at a position with lower sensitivity. For the A1 arc, we obtained
F$_{A1}\,=\,3.7_{-1.4}^{+5.1} \times 10^{-6} \, \rm erg \,  cm^{-2} \, s^{-1}$
with the same photon index. We consider the $F_{Bc}$ value the best flux estimate for the B arc,
corresponding to a fluence $F'_{Bc}\,=\,8.9_{-3.0}^{+6.1} \times 10^{-8} \, \rm  erg  \, cm^{-2}$.
 Moreover, we evaluate the flux at this position also assuming a photon index of -1.6 (observed, e.g.,
during the prompt phase of GRB 090510), obtaining $F_{hBc}\,=\,4.6_{-2.0}^{+5.5} \times
10^{-6} \, \rm erg \,  cm^{-2} \, s^{-1}$ and $F'_{hBc}\,=\,1.5_{-0.6}^{+1.8} \times\,
10^{-7} \, \rm  erg  \, cm^{-2}$.

We obtained the fluence extrapolations to the SA energy band to be
$\sim\,2.1\,\times\,10^{-8}\,\rm erg\, cm^{-2}$ for photon index of -2 and
$\sim\,4.8\,\times\,10^{-9}\,\rm erg\, cm^{-2}$ for -1.6.
We note that the SA UL fluence at the center of its FoV, $F^{SAc}$, is slightly lower
than the former value, therefore providing a constraint on the spectrum. The latter value of the extrapolated
fluence for the index of -1.6 is not constrained by the SA observations.
We also checked the extrapolation of the E2 flux to the GRID energy band and compared it
to the 2 s UL fluence. We obtained an extrapolated fluence at energies above 50 MeV of
$\sim\,8.9\,\times\,10^{-8}\,\rm erg\, cm^{-2}$  for a -2 photon index. Again, we end up
with a lower value with respect to the GRID 2 s UL fluence, $5.8\,\times\,10^{-6}\,\rm erg\, cm^{-2}$.}

{\fvv In conclusion, for} the \gw's most likely distance of 880 Mpc \cite[][]{2017PhRvLsub}, we obtain {\fvv for the E2 event in the MCAL band} an isotropic luminosity
$L_{\rm iso} = 2.6 \times 10^{50} \, \rm erg \,  s^{-1}$, and total
energy $E_{\rm iso} = 8.3 \times 10^{48} \, \rm erg$.

It is interesting to note that the E2 event
appears to be quite similar in its timing and flux characteristics
compared to the precursor event detected about 0.46 s before the
onset of the brightest part of the short GRB 090510. This weak
precursor was simultaneously detected by the AGILE-MCAL and
Fermi-GBM \cite[][]{2009Natur.462..331A,2010ApJ...708L..84G}, and
was therefore confirmed by the simultaneous detection of two space
instruments. In our case, this precursor of GRB 090510 played an
important role in calibrating the new MCAL data acquisition system
that has been optimized just for GW counterpart searches. {\fv We
notice that both the triggering events E1 and E2 (which occurred
within the telemetered time window after E1) satisfy} flux and
timing conditions set up following the MCAL detection of the GRB
090510 precursor\footnote[5]{The MCAL triggering procedure allows
only one trigger within the allocated time window following a
first event. The E2 event could have independently triggered the
MCAL system, but it did not because it was too close to E1.}. The
E2 event appears to be harder than the 2009 precursor.

If 
{\mt associated} with \gw, the E2 event shows interesting physical
properties. Its fluence and $L_{\rm iso}$ are comparable to those
of weak short GRBs previously detected by several space
instruments. A total energy estimated in the isotropic
approximation of $\simeq 10^{49} \, \rm erg$ is $10^{-7}$ times
smaller than the total black hole rest mass of $50 \, M_{\odot} \,
c^2$. Beaming effects would make this ratio even smaller. It is
clear that, if any association between E2 and \gw were confirmed
by simultaneous detections of different instruments, the e.m.
output following the \gw gravitational coalescence would be
severely constrained. MCAL cannot determine the position of the
event in the absence of simultaneous onboard detection by the
GRID or SA. However, from Fig. 1 we obtain the {\mt definite} sky
region belonging to the B arc that was not occulted by the Earth
and exposed by MCAL {\fvv for off-axis angles up to 45$^{\circ}$} at the $T_0$ of \gwp.
{\fvv We also have large off-axis angle MCAL exposures for regions labeled
A1 and A2 in Fig.~1.}

\section{Conclusions}

The LVC detection of the gravitational wave  \gw produced by a
BH--BH coalescence of a total mass of $50 \, M_{\odot}$ is a
remarkable event, the third in the LVC series of confirmed
detections. AGILE data are useful in 
{\mt assessing} the possible e.m. emission associated with GW
events. AGILE already obtained interesting upper limits for
gamma-ray and hard X-ray emission from the first GW event,
GW150914 \cite[][]{2016ApJ...825L...4T}. In that case, the
AGILE-GRID exposure of the LVC LR occurred 100 s before and 300
s after the event $T_0$.

 \gw is much more interesting from the point of view of AGILE
observations. In this case, significant exposure of the LR was
obtained simultaneously with the detection time $T_0$. The gamma-ray
and hard X-ray upper limits obtained by the GRID and SA are
therefore even more constraining than for GW150914.

Even more interestingly, we identify in the MCAL data the E2
candidate event with noticeable characteristics. We note its
similarity in flux and timing properties to the precursor of the
short GRB 090510. Our estimate of the post-trial probability of
being temporally associated with \gw (${\fvr 2.4}\,$--$\,2.7 \, \sigma$).
 As in the case of the GBR
090510 precursor, a confirmation by different space instruments
would provide the definite proof of {\fv its association with the GW}.

AGILE continues its observations of the sky in spinning mode and
is fully operational in the search of GW source counterparts.

\vskip .4cm \acknowledgements
We thank the LIGO/Virgo Collaboration for sharing information on \gw and for discussion and
comments on our manuscript. In particular, we thank F. Ricci, M. Branchesi, L. Singer, and
E. Katsavounidis. {\fvr We thank the anonymous referee for his/her comments.}
We thank the ASI management, the technical staff at the ASI Malindi ground station, the technical
support team at the ASI Space Science Data Center, and the Fucino AGILE Mission Operation Center.
 AGILE is an ASI space mission developed with programmatic support by INAF and INFN. We
acknowledge partial support through the ASI grant no. I/028/12/2.

\end{document}